\documentclass[aps,prb,twocolumn,showpacs,floatfix,groupedaddress]{revtex4-1}

\usepackage[utf8]{inputenc}
\usepackage{amsmath}
\usepackage{amsfonts}
\usepackage{graphicx}
\usepackage{bm}
\usepackage{float}
\usepackage{mathtools}
\usepackage{verbatim}
\usepackage{flushend}

\usepackage{booktabs}

\usepackage{bibentry}
\usepackage{natbib}

\usepackage{hyperref}

\newcommand{\eg}{\textit{e.g.}\ }
\newcommand{\ie}{\textit{i.e.}\ }

\begin{document}

\title{Infinite Matrix Product States vs Infinite Projected Entangled-Pair States on the Cylinder: a comparative study}

\author{Juan Osorio Iregui}
\email[]{osorio@itp.phys.ethz.ch}
\affiliation{Theoretische Physik, ETH Zürich, 8093 Zürich, Switzerland.}

\author{Matthias Troyer}
\email[]{troyer@phys.ethz.ch}
\affiliation{Quantum Architectures and Computation Group, Microsoft Research, Redmond, WA 98052, USA.}

\author{Philippe Corboz}
\email[]{P.R.Corboz@uva.nl}
\affiliation{Institute for Theoretical Physics, University of Amsterdam\\
Science park 904 Postbus 94485, 1090 GL Amsterdam, The Netherlands.}

\date{\today}

\begin{abstract}
	In spite of their intrinsic one-dimensional nature matrix product states have been systematically used to obtain remarkably accurate results for two-dimensional systems. Motivated by basic entropic arguments favoring projected entangled-pair states as the method of choice, we assess the relative performance of infinite matrix product states and infinite projected entangled-pair states on cylindrical geometries. By considering the Heisenberg and half-filled Hubbard models on the square lattice as our benchmark cases, we evaluate their variational energies as a function of both bond dimension as well as cylinder width. In both examples we find crossovers at moderate cylinder widths, \ie for the largest bond dimensions considered we find an improvement on the variational energies for the Heisenberg model by using projected entangled-pair states at a width of about 11 sites, whereas for the half-filled Hubbard model this crossover occurs at about 7 sites.
\end{abstract}


\maketitle

\section{Introduction}

One of the main forces driving the area of tensor network algorithms (TNAs) stems from the remarkable success of the so-called density matrix renormalization group (DMRG) algorithm\cite{white1992density} in the simulation of 1d lattice models. This success has, for a long time now, been understood as a consequence of a very specific property of ground states of 1d local hamiltonians. As shown in a beautiful piece of work by Hastings,\cite{hastings2007entropy} the combined effect of local interactions and an energy gap for excitations results in the remarkable property that ground states of such hamiltonians both obey what has been dubbed as an \emph{area law} of entanglement entropy, \ie the property that the von Neumann entropy of the system upon bipartition only scales proportionally to the surface of the boundary connecting both subsystems, as well as admit an efficient representation in terms of so-called matrix product states (MPS). A class of states obeying a 1d area law and which precisely makes up the manifold of states over which the DMRG operates.

Rather remarkable is also the fact that the class of MPS has been, repeatedly, found to provide very accurate results even when studying \eg critical systems,\cite{tagliacozzo2008,pollmann2009theory,pirvu2012matrix,mong2014parafermionic,stojevic2015conformal} in which the closing of the energy gap leads to a spreading of correlations over all length scales in the system, or 2d systems of considerable dimensions,\cite{depenbrock2012nature,kolley2015phase,varjas2013chiral,leblanc2015solutions} where an area law can no longer be used to certify the efficiency of MPS. As a consequence MPS have become not only the golden standard for the simulation of 1d lattice models, but also one of the most competitive algorithms for the simulation of strongly correlated 2d systems.

In the realm of two-dimensional TNAs developments generalizing MPS to the higher-dimensional setting have been targeted at the adequate incorporation of the amount of entanglement expected for gapped 2d phases, with one of the primary examples being projected entangled-pair states (PEPS),~\cite{verstraete2004renormalization,verstraete2006criticality} also called tensor product states,~\cite{nishino2001,nishio2004} a class of ansatz states which, by construction, satisfy a 2d area law.

Even though significantly younger than MPS, PEPS have already shown considerable promise as a competitive algorithm for the simulation of strongly correlated systems, with notable achievements including some of the lowest variational energies along with important insights into the physics of the $t$-$J$,~\cite{corboz2014competing} Hubbard,~\cite{corboz16} as well as numerous frustrated spin models (see \eg Refs.~\onlinecite{corboz11-su4,wang11_j1j2,Corboz12_su4,corboz2013tensor,xie14,corboz2014crystals,iregui2014probing,picot15,picot15a,nataf16,liao2016gapless,niesen2017emergent} and references therein) in the thermodynamic limit, where the algorithms acquire the name of infinite PEPS.\cite{jordan2008classical}

Given the intrinsic difference regarding the entanglement scaling that each of these algorithms is designed to support, a few questions naturally begin to arise, namely: is it possible to employ PEPS wavefunctions as a complementary approach once entropic demands start becoming prohibitively large for MPS simulations on 2d systems? If so, for which system sizes should this takeover begin to happen?

By considering a couple of paradigmatic benchmark cases, \ie the Heisenberg and half-filled Hubbard models on the square lattice, we will begin to address this question here and argue that using PEPS on infinite cylinders can indeed provide a powerful complementary approach to MPS already at moderate widths.

This paper is structured as follows: in section \ref{sec:methods} we present brief summaries of the (infinite) MPS (iMPS) and (infinite) PEPS (iPEPS) algorithms. Sec. \ref{sec:motivation} gives the main motivation behind this work. We present our main results in section \ref{sec:results}, where we simulate the Heisenberg and half-filled Hubbard models on square lattice cylinders of various widths, and then conclude with a few remarks in section \ref{sec:discussion}.
Supporting information as well as numerical data can be found in the various appendices.

\section{Methods}\label{sec:methods}

\subsection{Matrix Product States}
\begin{figure}[h]
	\centering
	\includegraphics[trim={8cm 10cm 0cm 0cm},clip,width=0.9\textwidth]{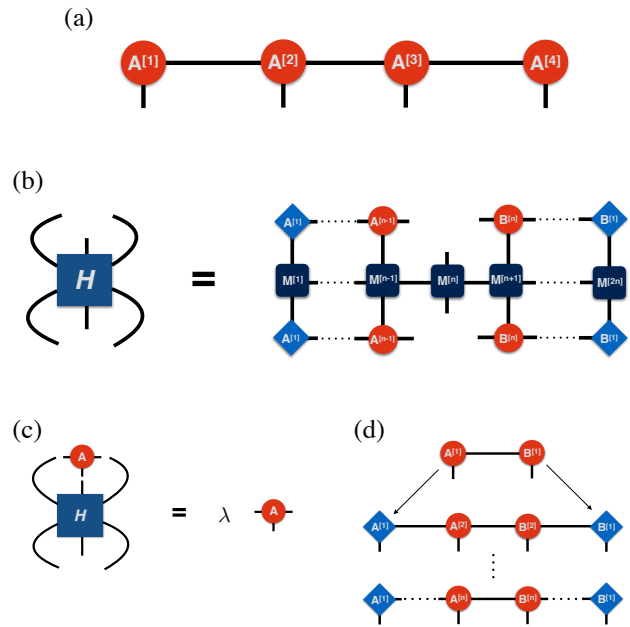}
	\caption{Summary of iMPS set-up. (a) Four-site MPS ansatz wavefunction. (b) Definition of the effective single-site hamiltonian $H$ in Eq.~\ref{eq:regular-eigenproblem}, square tensors labelled by M in the middle row are used to give a so-called \emph{matrix product operator} representation of the hamiltonian $\mathcal{H}$ for the full system. (c) Graphical representation of eigenvalue problem in Eq.~\ref{eq:regular-eigenproblem}. (d) iMPS growing procedure, where red tensors (circles) in the middle are optimized whereas blue tensors (diamonds) correspond to tensors from previous iterations, which are no longer accessible for optimization.}\label{fig:mps_summary}
\end{figure}

The class of MPS can be defined on a $N$-component system, here taken to be spins for simplicity, as the manifold of states $\lvert \psi \rangle$ obtained as

\begin{align}
	\lvert \psi \rangle = \sum_{\{S\}} Tr[ A^{[1]}_{S_1} A^{[2]}_{S_2} \cdots A^{[N]}_{S_N} ] \lvert S_1,S_2,\cdots,S_N \rangle,\label{eq:mps_ansatz}
\end{align}

where each of the $A^{[i]}$ represents a rank-3 tensor of dimensions $[m^{[i]},m^{[i+1]},|S|]$, see Fig.~\ref{fig:mps_summary}(a); the last dimension is fixed by the physical degree of freedom and $m^{[i]}$, $m^{[i+1]}$, represent the number of rows and columns, respectively, of the matrix $A^{[i]}_{S_i}$ obtained by fixing a physical state $S_i$. The fact that the amplitude $\psi_{S_1,S_2,\cdots,S_N}$, corresponding to a basis state $\lvert S_1,S_2,\cdots,S_N \rangle$, is given by the trace over a product of matrices motivates the name of the ansatz. Moreover, it is customary to refer to the largest dimension of the matrices making up the ansatz, \ie the largest $m^{[i]}$, as the \emph{bond dimension}, which we will hereafter denote as $m$.

In practice the entries of the matrices defining the ansatz are used as a set of variational parameters to be optimized in order to minimize the energy of the state. This is most often done by minimizing the cost functional given by the equation

\begin{equation}
	f[\{A\}] = \langle \psi \rvert \mathcal{H} \lvert \psi \rangle + \lambda (1 - \langle \psi \lvert \psi \rangle ) \label{eq:mps_cost_function}
\end{equation}

using a procedure, known as the DMRG algorithm, in which the entries of each tensor $A$ are optimized in a sweeping pattern, one tensor at a time, until a minimum is obtained. Since the class of MPS provides a variational ansatz, $f[A]$ represents a functional with its minima at each of the ground states of hamiltonian $\mathcal{H}$, in which case the constant $\lambda$ will represent the ground state energy $E_0$. By using a so-called canonical form\cite{schollwock2011density} the optimal set of entries for a target tensor may be obtained by solving the eigenvalue problem

\begin{equation}
	H \mathbf{a} = \lambda \mathbf{a} \label{eq:regular-eigenproblem}
\end{equation}

defined in Fig.~\ref{fig:mps_summary}(c), where $\mathbf{a}$ represents a vectorization of the target tensor and $H$ may be regarded as an effective single-site hamiltonian. By iterating this procedure through all tensors in the ansatz the energy of the state can be minimized to obtain, in general, an approximation to the true ground state. Such a procedure can be carried out with a total memory cost of $O(m^2)$ and a computational complexity of $O(m^3)$.

\subsection{infinite Matrix Product States}

The DMRG algorithm above can also be generalized to simulate systems of increasing size, thus acquiring the name iDMRG. Indeed, an infinite MPS simulation can be formulated as a regular MPS optimization which only has access to a given subset of tensors in the wavefunction. As is illustrated in Fig.~\ref{fig:mps_summary}(d), where we use a 2-tensor growing pattern for simplicity, once the initial two-site problem has converged one may use these tensors as the external (blue tensors) part of a new system consisting of 2 + 2 sites, for which only the two central tensors can be optimized. By continuing this pattern one ends up with a $2s$-site system after $s$ iterations have concluded. Once this procedure converges, the central sites can be used to represent a unit cell of an infinite-size system, from which one may readily evaluate observables in the thermodynamic limit.

In the case of 2d MPS simulations, it is typical to consider a cylindrical geometry: from a physical perspective, the cylinder allows to preserve the translation symmetry along one of the axes while at the same time minimizing finite size effects; this argument would appear to favor toroidal geometries, however, from the numerical side, keeping open boundary conditions along one lattice dimension naturally allows to do the same with the MPS ansatz, preserving the stability of the algorithm without imposing additional entanglement requirements on the wavefunction.

Importantly, both the memory and computational complexities above remain the same regardless of the geometry of the system being simulated, albeit at the expense of simulating a system with longer-ranged interactions.

For a more detailed discussion of the DMRG and iDMRG algorithms we refer the interested reader to the excellent reviews in Refs. [\onlinecite{schollwock2011density,kjall2013phase,zaletel2015infinite}].

\subsection{Projected Entangled-Pair States}
\begin{figure}[h]
	\centering
	\includegraphics[trim={8cm 9cm 0cm 0cm},clip,width=0.85\textwidth]{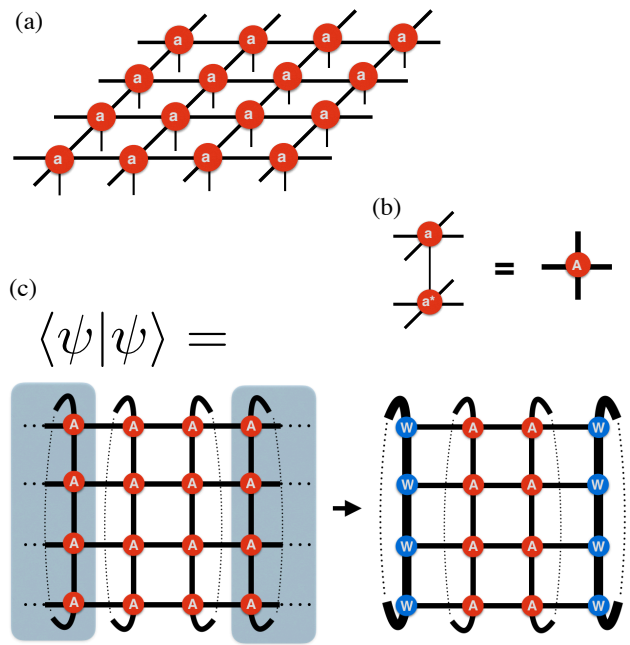}
	\caption{Summary of iPEPS set-up on the cylinder. (a) iPEPS ansatz wavefunction with a single-site unit cell, (b) Definition of the double layer tensor $A$ corresponding to the contraction of unit cell tensor $a$ and its conjugate, (c) tensor network representing the square of the norm of an iPEPS state $\lvert \psi \rangle $ on an infinite cylinder, illustrating the effective representation of semi-infinite environments as boundary tensors W.}\label{fig:peps_summary}
\end{figure}

The PEPS ansatz wavefunction is defined in a manner analogous to Eq.~\eqref{eq:mps_ansatz}, by generalizing it to

\begin{align}
	\lvert \psi \rangle = \sum_{\{S\}} tTr[ A^{[1]}_{S_1} A^{[2]}_{S_2} \cdots A^{[N]}_{S_N}] \lvert S_1,S_2,\cdots,S_N \rangle, \label{eq:peps_ansatz}
\end{align}

where each tensor $A$ now represents a rank-($h$+1) tensor, with $h$ the coordination number of the lattice, and $tTr$ represents the tensor trace, \ie a summation over all virtual (auxiliary) indices, see Fig.~\ref{fig:peps_summary}(a). The bond dimension here refers to the size of the largest virtual index and is traditionally denoted by $D$. As in the MPS case the entries of tensors $A$ are regarded as variational parameters to be optimized in order to minimize the energy of the system.

In practice, carrying out the exact summation over virtual indices in Eq.~\eqref{eq:peps_ansatz}, as required \eg whenever computing expectation values of observables, generically involves a computational cost scaling exponentially with the smallest system dimension. It is therefore necessary to introduce so-called \emph{contraction schemes}, \ie approximations of the tensor trace in Eq.~\eqref{eq:peps_ansatz}. The computational cost of contraction schemes used in practice is typically located in the range $O(D^{10})-O(D^{12})$.\cite{nishino1996corner, nishino97,orus2009simulation,verstraete2004renormalization,jordan2008classical,lubasch2014algorithms,lubasch2014unifying,corboz2014competing,zhao2010renormalization,zhao2016tensor}

In the case of PEPS (and iPEPS) the optimization of the tensors has often been carried out using so-called \emph{imaginary-time evolution} (ITE), in which an initial state $\lvert \psi \rangle$ is projected onto the ground state by acting with the operator $\exp\left( -\beta \mathcal{H} \right)$, \ie

\begin{equation}
	\lvert \psi_0 \rangle = \lim_{\beta\to\infty} \exp\left( -\beta \mathcal{H} \right) \lvert \psi \rangle.
\end{equation}

Given the considerable computational cost of optimizing PEPS wavefunctions, additional approximations are typically used in practice. The simplest, and also less-accurate, of these receives the name of \emph{simple update}\cite{jiang2008accurate} (SU) as it involves only the pair of PEPS tensors directly affected by a single local ITE gate. An improvement upon this corresponds to the so-called \emph{full update}\cite{jordan2008classical,lubasch2014algorithms,phien2015infinite} (FU) in which each optimization is still carried out only with respect to a single ITE gate yet the full wavefunction is taken into account. Finally, the most accurate scheme relies on the minimization of Eq.~(\ref{eq:mps_cost_function}) as in the MPS case. This scheme is refered to as the \emph{variational update} (VU).\cite{nishino2001,gendiar03,verstraete2004renormalization,Verstraete08,vanderstraeten2016gradient,corboz2016variational} In this work we present data obtained from all three schemes, depending on the model under consideration.

Once all approximations have been introduced, the PEPS algorithm can be carried out with a total memory cost of $O(D^h)$ and a leading computational complexity of $O(D^{10})-O(D^{12})$.

\subsection{infinite Projected Entangled-Pair States}
It is possible to generalize the PEPS ansatz to infinite systems by choosing unit cells of tensors which are periodically repeated in all directions. For instance, the case of a single-site unit cell is illustrated in Fig.~\ref{fig:peps_summary}(a).

One of the useful properties of (i)PEPS wavefunctions is their modularity. Given that once a particular unit cell structure has been chosen and PEPS tensors representing states of interest have been obtained, \ie either by numerical optimization or by explicit construction, these may be embedded in any lattice admiting a complete covering based on the PEPS unit cell. Thus allowing to readily obtain ansatz states for a range of lattice sizes directly from a single simulation. This is a feature that has already been exploited before\cite{wang2011monte,wang2013constructing} and that we will take advantage of for all results below, \ie all results presented here have been obtained by using PEPS tensors optimized on infinite lattices.

The case of computations on infinite cylindrical geometries, see Fig.~\ref{fig:peps_summary}, has received reduced attention,\cite{ladderspecializations} with a lot of the work focused on the extraction of topological information\cite{cirac2011entanglement,poilblanc2015chiral,schuch2013topological} in which case the computations may be carried out using cylinders of infinite width, which allow for convenient simplifications, or using exact contractions of cylinders with modest widths. The case of numerical simulations on cylinders with an \emph{arbitrary} finite width has received close to no attention. Since, in order to obtain the results we present here, a large part of the work involved the development and testing of various contraction schemes targeted towards cylindrical geometries, we will postpone a detailed presentation of the techniques employed to a follow-up publication.\cite{osorioireguicylinderipeps}

For the sake of completeness we have nevertheless included a very brief summary of a procedure used for the iPEPS simulations in Appendix \ref{appendix:ipeps}.

\section{Motivation}\label{sec:motivation}
Even though numerous studies have made it clear that the class of MPS provides a very competitive approach for the simulation of strongly correlated systems, even beyond 1d, the fact that generic ground states of systems in higher dimensions impose heavier requirements on the amount of entanglement an ansatz wavefunction must be able to encode, poses serious questions to the scalability of MPS as the method of choice for their study.

\begin{figure}
	\centering
	\begin{tabular}{ll}
		\includegraphics[trim={4.5cm 2cm 5cm 2cm},clip,width=0.24\textwidth]{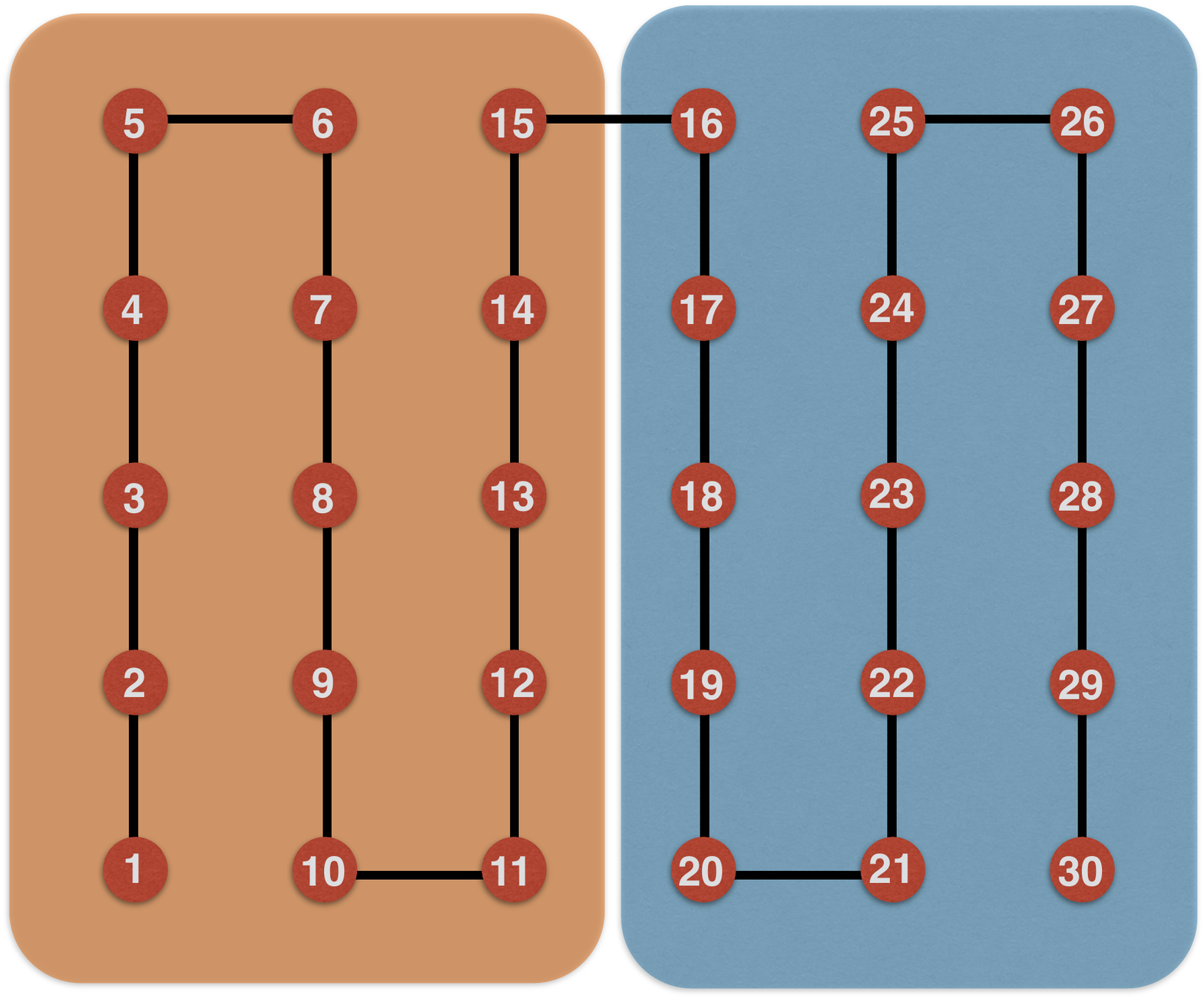}
		&
		\includegraphics[trim={5cm 2cm 4.5cm 2cm},clip,width=0.24\textwidth]{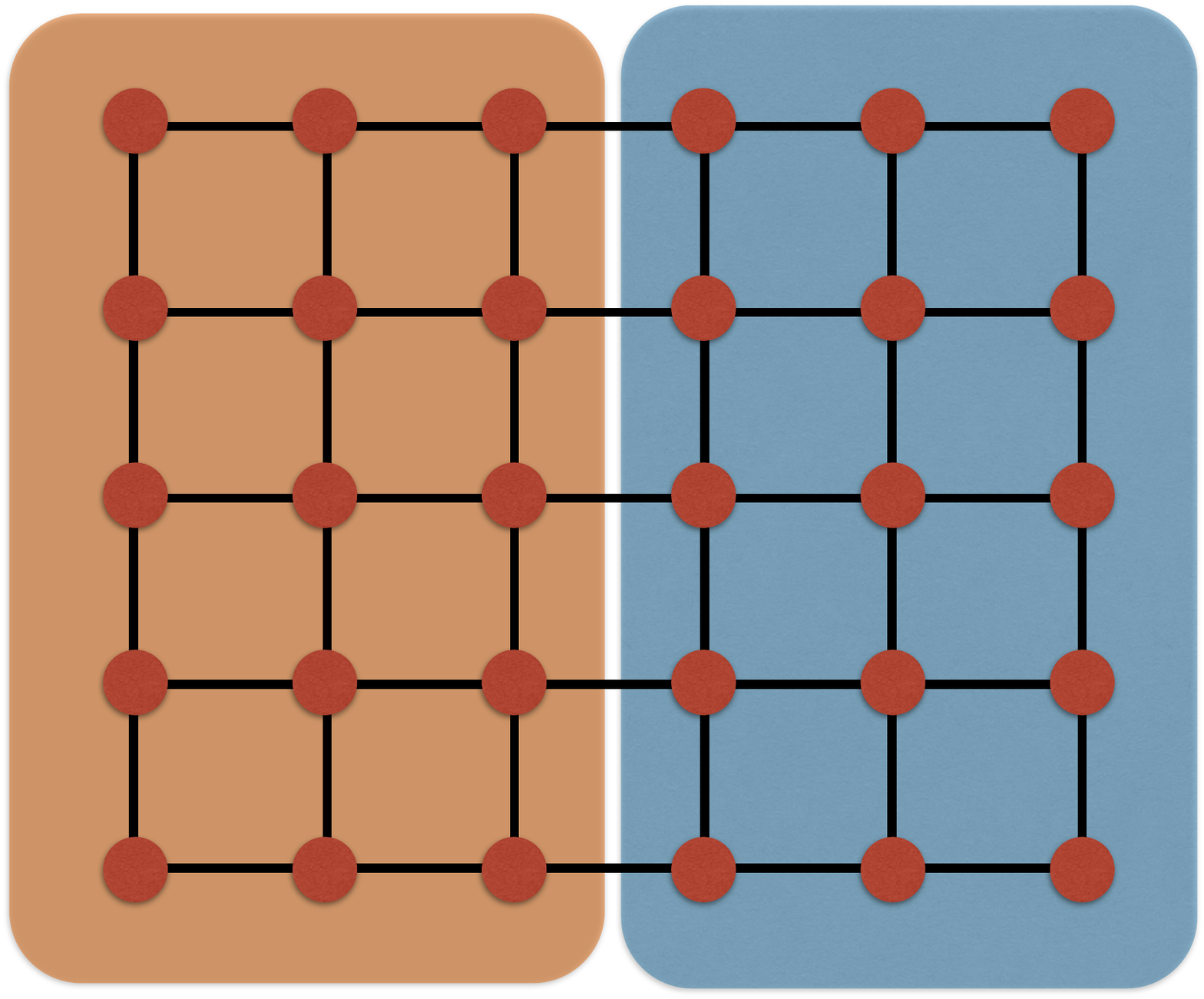}
	\end{tabular}
	\caption{Possible coverings for the simulation of 2d systems using MPS (left) and PEPS (right) on a square lattice. The brown and blue halves illustrate a possible bipartition of the system.}
	\label{fig:mps_vs_peps_mapping}
\end{figure}

Let us consider a scenario in which the ground state of a 2d system is known to obey an area law. One possible way of encoding this wavefunction would be to map it onto a, intrinsically 1d, MPS wavefunction by numbering sites on the 2d lattice sequentially, \eg in a snake pattern as in Fig.~\ref{fig:mps_vs_peps_mapping}. Another approach would correspond to using a PEPS wavefunction, with which the connectivity of the lattice would be naturally reproduced, see Fig.~\ref{fig:mps_vs_peps_mapping}.

Upon bipartition of the system as \eg in Fig.~\ref{fig:mps_vs_peps_mapping}, the entanglement entropy $S_{MPS}$ and $S_{PEPS}$ of the ansatz states will be upper bounded by their corresponding bond dimension as
\begin{align*}
	S_{MPS} \sim& \log(m),\\
	S_{PEPS} \sim& W \log(D),
\end{align*}

with $W$ the width of the system, or the (interface) surface area of the bipartition. Since both states are supposed to encode the same underlying wavefunction, naively equating entropies gives

\begin{equation}
	m \sim D^W.
\end{equation}

In other words, the bond dimension required for the MPS encoding would grow exponentially fast with the width of the system compared to the bond dimension required for the PEPS. This can be seen as a reflection of the fact that, whenever dealing with 2d systems, MPS are known to require an exponential scaling of bond dimension in the smallest system dimension in order to preserve a given accuracy.\cite{liang1994approximate}

To get an idea of the actual numbers one might consider a value of $D = 4$, which is well within the limits of what is currently reachable in a numerical simulation, and a width of $W=10$, which gives $m = 1.048.576$, a value vastly exceeding what is currently feasible using state-of-the-art implementations of the DMRG algorithm. Thus, in light of this simple entropic argument, it might appear obvious what the method of choice should be whenever simulating 2d systems. However, as we have remarked in the previous section, PEPS simulations do not inherit certain advantageous features, related to numerical stability and computational complexity present in MPS simulations, \eg the difference in computational scaling $O(m^3)$ vs. $O(D^{10})$ allowing for substantially larger value of $m$, making it difficult to predict their relative performance based on such a naive argument.

\section{Results} \label{sec:results}

\subsection{Heisenberg model}
In this section we will consider the Heisenberg model

\begin{equation*}
	\mathcal{H} = \sum_{\langle i , j \rangle} \hat{S}_i \cdot \hat{S}_j,
\end{equation*}
where, here and below, $\langle i , j \rangle$ represents a sum running over all nearest-neighbor bonds of a square lattice cylinder, \ie using periodic boundary conditions along the vertical direction and open boundaries along the horizontal direction, of width $W$ and $\hat{S}$ represents a $S=1/2$ spin operator. For $W=2$ cylinders the system becomes strongly dimerized, with a finite spin gap separating the ground state from a band of propagating spin triplets. Even though this picture remains valid for any finite even width, increasing the system width has the effect of reducing the energy gap. In the limit $W\rightarrow\infty$ the energy gap closes, giving way to a critical state with an algebraic decay of correlations and a finite sublattice magnetization. The most accurate results to date,\cite{sandvik2010computational} using stochastic series expansions (SSE), give an energy per site $E_0 = -0.6694421(4)$ in the thermodynamic limit.

As this model is free of frustration for even widths, we will constrain ourselves to such systems for which we will use loop quantum Monte Carlo (QMC) simulations to obtain reference data. See Appendix \ref{appendix:lqmc} for further details and Appendix \ref{appendix:numerical_data} for a table containing our loop QMC energy estimates.

\begin{figure}[!h]
	\hspace*{-0.29cm}
	\includegraphics[width=0.5\textwidth]{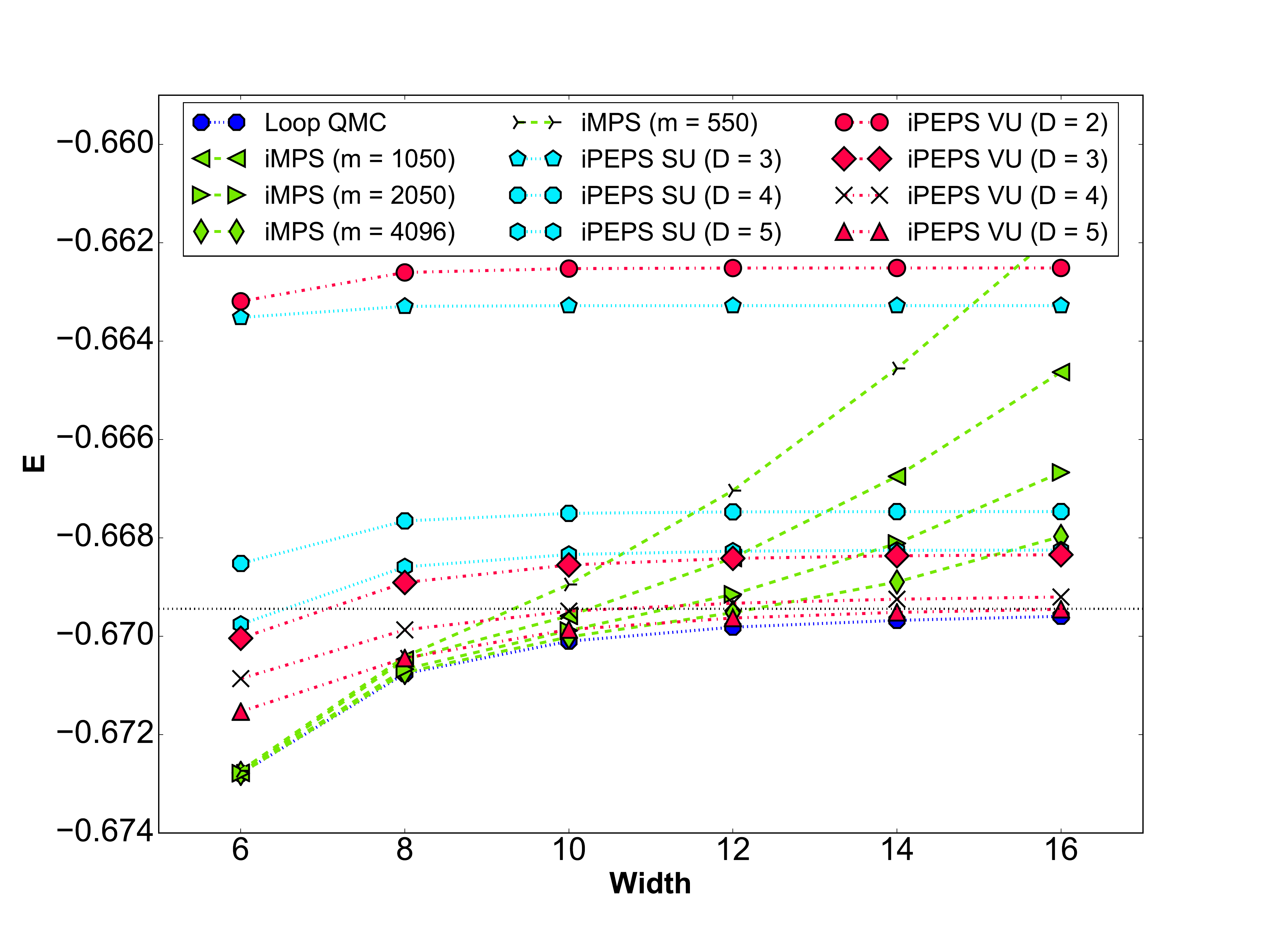}
	\caption{Variational energies for the Heisenberg model using iMPS (dashed lines/green symbols), iPEPS SU (dotted lines/cyan symbols), iPEPS VU (dot-dashed lines/red symbols) and loop QMC (dotted line/blue circles) methods. Loop QMC error bars are smaller than the symbol size. The black horizontal dotted line represents the SSE thermodynamic limit estimate, $E_0=-0.6694421(4)$, from Ref.~[\onlinecite{sandvik2010computational}]. For the largest bond dimensions considered, \ie $m=4096$ and $D=5$, a crossover between iMPS and iPEPS is visible around $W\sim11$.}
	\label{fig:heisenberg-imps-vs-ipeps-width}
\end{figure}

Our main results for the Heisenberg model are shown in Fig.~\ref{fig:heisenberg-imps-vs-ipeps-width}. As initially expected, various crossings of the energy curves appear depending on the precise values of the bond dimension as well as optimization scheme. For the largest bond dimensions considered, \ie $m=4096$ and $D=5$, the crossover where iPEPS outperform iMPS happens at a width $W \sim 11$, although a very close competition is clearly visible in the range $W \in [8,12]$.

For the iPEPS simulations we find that, up to the largest bond dimension considered ($D=5$), most curves corresponding to the SU remain remarkably flat across the full set of widths simulated. Given that the SU is based on an approximation incorporating only subparts of the wavefunction, renormalization effects arising due to longer-ranged entanglement are almost completely absent and, in this case, it simply produces states for which the correlation length does not appear to become large enough to notice the finite width of most cylinders considered. Indeed, the values of the energy obtained contain only minor corrections to the value obtained for the 2d system.

\begin{figure}[!htbp]
	\hspace*{-0.37cm}
	\includegraphics[scale=0.235]{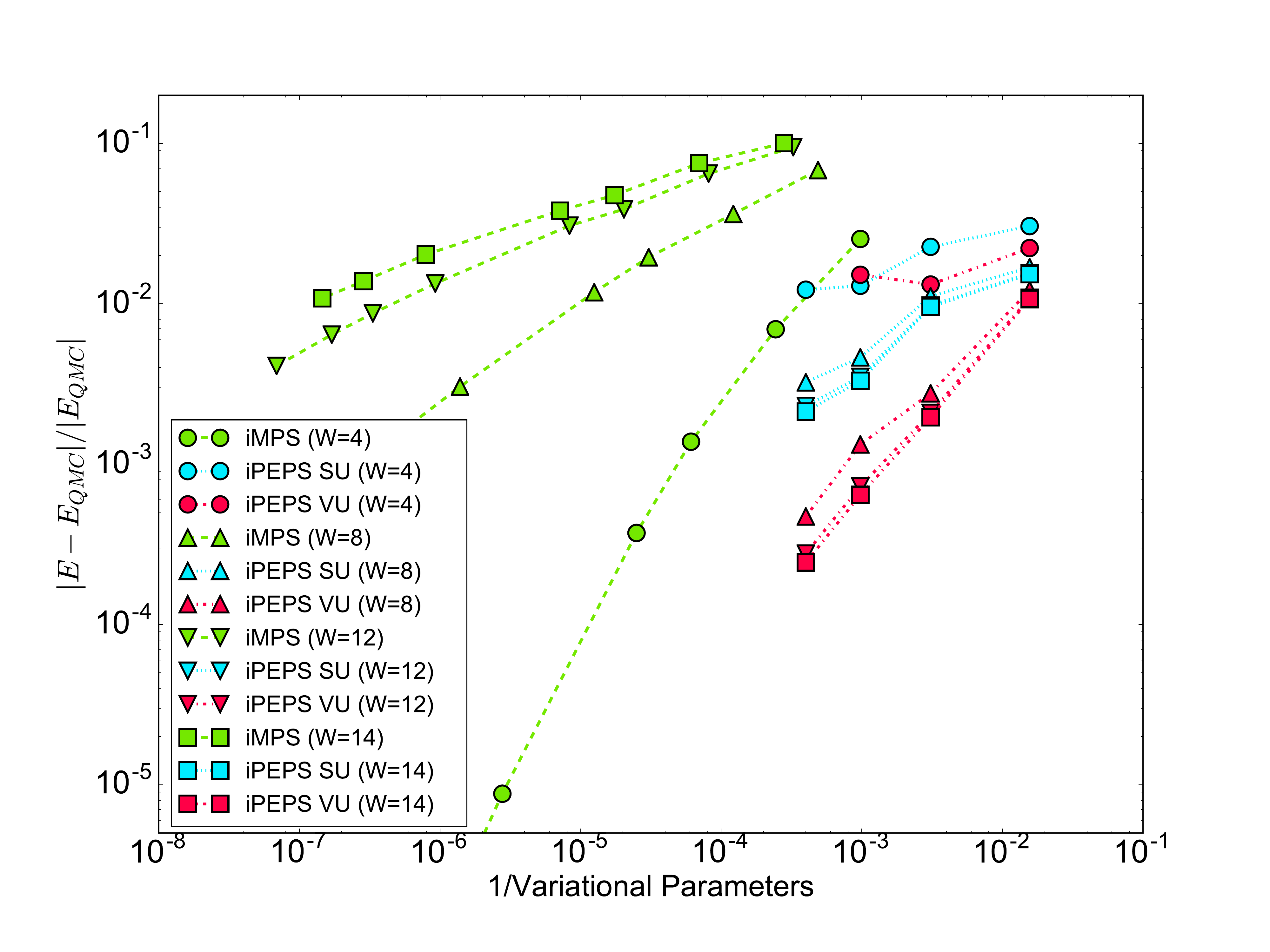}
	\caption{Relative error of the different ansatz wavefunctions for the square lattice Heisenberg model as a function of inverse variational parameters. The higher efficiency of the iPEPS ansatz (SU - dotted lines/cyan symbols, VU - dot-dashed lines/red symbols) is reflected on the fact that all iPEPS curves are located to the right of all iMPS curves (dashed lines/green symbols). }
	\label{fig:heisenberg-imps-vs-ipeps-var-params}
\end{figure}

On the other hand simulations with tensors using the VU improve significantly on the energies of the SU and a bending of the energy curves is clearly visible as one moves to narrower cylinders, although not enough to completely match the accuracy of the iMPS simulations for the narrowest cylinders shown ($W=6$). The overall improvement of the VU compared to the SU can be understood as a consequence of the tensors providing more accurate approximations to the physics of the infinite size system. Still, given that we are ultimately not properly accounting for finite size effects, as one moves to narrower cylinders it can clearly be seen that the relative accuracy drops considerably below $W\sim8$.

We find that iMPS simulations manage to reproduce the reference energy values to high accuracy for cylinder widths up to $W \sim 10$ where, for a given bond dimension $m$, a clear up-bend in the curves starts to take place resulting in a significant loss in accuracy. This is a clear reflection of the fact that the bond dimensions considered are not enough to compensate for the higher entropic demands of simulations on wider cylinders.

To get an overall idea of how efficient the encoding of the wavefunctions is, in Fig.~\ref{fig:heisenberg-imps-vs-ipeps-var-params} we show a comparison of the relative errors for both iMPS and iPEPS as a function of inverse number of variational parameters. There it is again possible to see how, for a fixed number of variational parameters, the VU provides a significantly more accurate estimate than that of the SU. More importantly, it is also possible to see how the VU curves systematically decay faster than the SU curves as one increases the number of variational parameters. It is remarkable that, starting with cylinders of width $W \sim 8$, the rate of decrease in relative error for iPEPS wavefunctions as a function of inverse number of variational parameters essentially matches that of iMPS on the narrowest cylinders considered ($W=4$). This observation becomes all the more relevant once we recall that these tensors have not been optimized for each of the cylinder widths.

We believe these are very promising results as, already at this intermediate step, it seems like a small increase in bond dimension for iPEPS should yield data comparable to our reference QMC data, for a wide range of cylinder widths.

\subsection{Hubbard model}
We now  consider the Hubbard model
\begin{equation*}
	\mathcal{H} = - t \sum_{\langle i , j \rangle,\sigma} \left( \hat{c}^\dagger_{i\sigma} \hat{c}_{j\sigma} + h.c. \right) + U \sum_{i} \hat{n}_{i\uparrow} \hat{n}_{i\downarrow},
\end{equation*}
where $\hat{c}^\dagger_{i\sigma}$ ($\hat{c}_{i\sigma}$) creates (annihilates) an electron with spin $\sigma$ on site $i$ and $\hat{n}_{i\sigma} := \hat{c}^\dagger_{i\sigma} \hat{c}_{i\sigma}$ represents the number operator. This model has been studied extensibly using a large variety of numerical methods given its close connection to the physics of the cuprate high-temperature superconductors. In the half-filled case, \ie $ n := \frac{1}{N} \sum_i \langle n_i \rangle = 1 $ and $N$ the number of sites, it is widely accepted that the system finds itself in a Mott-insulating regime for arbitrarily small values of the on-site repulsion $U$. Since we wish to avoid difficulties arising due to a large number of competing states, a problem largely present at weak doping, we shall constrain our simulations to the half-filled regime at a strong repulsion of $U/t=8$. Using this set of parameters the ground state energy per site has been estimated using auxiliary field QMC (AFQMC) to be $E_0=-0.5247(2)$ in the thermodynamic limit.\cite{leblanc2015solutions}

\begin{figure}[h]
	\hspace*{-0.29cm}
	\includegraphics[scale=0.235]{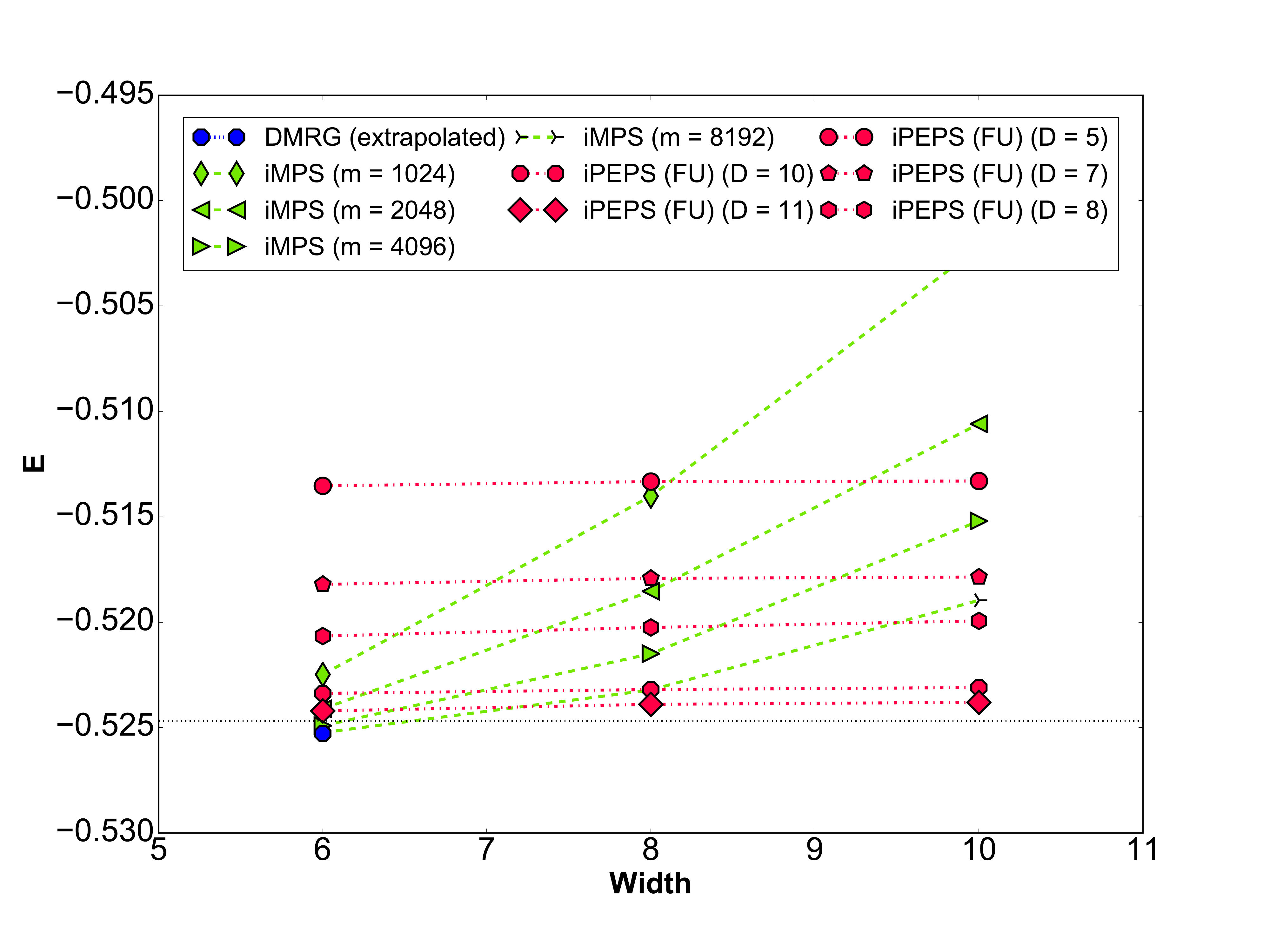}
	\caption{\label{fig:Hubbard-imps-vs-ipeps-width} Variational energies for the half-filled Hubbard model at $U/t=8$, using iMPS (dashed lines/green symbols), iPEPS (dot-dashed lines/red symbols) and (extrapolated) DMRG (blue circle) methods. The DMRG result was obtained from Ref.~[\onlinecite{leblanc2015solutions}]. The horizontal dotted line represents the AFQMC thermodynamic limit estimate, $E_0=-0.5247(2)$, from Ref.~[\onlinecite{leblanc2015solutions}]. For the largest bond dimensions considered, \ie $m=8192$ amd $D=11$, a crossover between iMPS and iPEPS is visible around $W\sim7$. }
\end{figure}

The ground state energy per site on an infinite cylinder of width 6 has been estimated using state-of-the-art finite DMRG simulations,\cite{leblanc2015solutions} employing ellaborate finite-size-effect cancellation and extrapolation techniques, to be at $E^{(6)}_0=-0.52528(1)$.

Our main results for the half-filled Hubbard model are summarized in Fig.~ \ref{fig:Hubbard-imps-vs-ipeps-width}. In general we find a situation qualitatively similar to that in Fig.~\ref{fig:heisenberg-imps-vs-ipeps-width} for the Heisenberg model, in that various energy crossings are visible depending on the different values of the bond dimensions $m$ and $D$. In this example we find that for the largest bond dimensions considered, \ie $D=11$ and $m=8192$, the crossover where iPEPS provides an improvement over iMPS happens at a width of $W\sim7$.

For our iMPS simulations we find good convergence at $W=6$, where we may directly compare to Ref.~[\onlinecite{leblanc2015solutions}], after which a strong increase in energy is noticeable for all bond dimensions considered. This more rapid increase in iMPS energies, compared to the one found above for the Heisenberg model, can be understood as a consequence of the increased local Hilbert space dimension, \ie 4 compared to 2 of the Heisenberg model, thus allowing for a more rapid build-up of entanglement between different parts of the system as the cylinder widths increase. Even though the largest value of $m$ we consider here, \ie $m=8192$, still does not quite manage to reproduce the (extrapolated) reference energy quoted above, it is nevertheless reassuring to see that the difference found is quite small, with a relative difference $\Delta E^{(6)} \approx 0.01\%$ for a cylinder with $W=6$. It is also interesting to see that without any additional extrapolation the energies of the largest bond dimension iPEPS simulations already come very close to the thermodynamic limit estimate, with a relative error of $\Delta E = 0.17\%$.

\section{Discussion} \label{sec:discussion}
We have managed to show how, as expected from a naive entropic argument, various crossovers in the relative accuracy of the iMPS and iPEPS ansätze occur as the width of the cylinders and the bond dimensions $m$ and $D$ are increased. For the largest bond dimensions considered, a modest width of $W\sim7$ was already enough to obtain an improvement over iMPS energies by using iPEPS wavefunctions when simulating the half-filled Hubbard model. On the other hand we found that when simulating the Heisenberg model this crossover takes place at a considerably larger width of $W\sim11$, albeit with a strong competition over a considerable range of widths.

It is perhaps worth emphasizing that it is not the \emph{precise} widths at which the crossings take place that are of particular importance but, instead, it is the overall energy landscape which provides important insights. After all it is clear that modifications in either bond dimension will necessarily lead to a shift of the crossover value (not to mention improvements in the algorithms). The energy landscape, however, gives us a qualitative understanding of their relative accuracies and a, admittedly rough, $m$-to-$D$ mapping under different settings which, together with knowledge of the actual computational effort for each simulation, can be used as a guiding principle for when to choose one method over the other.

Going back to our original estimate for the values of the bond dimensions expected, it is clear that the naive argument presented above resulted in a gross overestimation of the iMPS bond dimension required to achieve an accuracy comparable to iPEPS. Indeed, from the results we have presented, instead of $m=1.048.576$ expected to match the accuracy of the $D=4$ iPEPS ansatz at $W=10$, we found that $m \sim 1000$ was already enough to achieve comparable results for the Heisenberg model using iMPS.

We may take away a few important messages from this first exercise: first, the data presented here reminds us of the remarkable stability and accuracy of the (i)DMRG algorithm, allowing it to achieve very competitive results for considerable widths at moderate bond dimensions; second, it also shows that the room for improvement on the PEPS side is quite large. Given that, in spite of all the short-comings intrinsic to the iPEPS algorithm, we were able to show improved results at modest to intermediate cylinder widths without specifically optimizing the PEPS tensors for each cylinder, we expect to see additional improvements once this tuning is introduced, particularly at the small to intermediate width range.

We believe that these results provide a good example of how MPS and PEPS algorithms may be employed in the future in a complementary way to obtain accurate results over a wide range of cylinder widths, exploiting the remarkable accuracy of MPS at reduced cylinder widths while making full use of the entropic advantage provided by PEPS at increased widths, with the intermediate region serving as a direct cross-check scenario.

Motivated by the promising results that we have presented here, we are currently exploring different variants of both improved contraction and optimization schemes specifically targeted at cylindrical geometries, which we plan to make public in the near future.\cite{osorioireguicylinderipeps}

\begin{acknowledgements}

	J.O.I. would like to acknowledge useful discussions with Michele Dolfi and Guifré Vidal. Most simulations were performed using the Mönch cluster of the Platform for Advanced Scientific Computing (PASC) at ETH Zürich. This project has received funding from the European Research Council (ERC) under the European Unions Horizon 2020 research and innovation programme (grant agreement No 677061). This work is part of the D-ITP consortium, a program of the Netherlands Organisation for Scientific Research (NWO) that is funded by the Dutch Ministry of Education, Culture and Science (OCW). This work
	was also part of the FOR1807 Advanced Computational Methods for Strongly Correlated Quantum Systems research unit of the Deutsche Forschungsgemeinschaft (DFG).

\end{acknowledgements}

\appendix

\section{Loop Quantum Monte Carlo Simulations of the Heisenberg Model} \label{appendix:lqmc}
As was mentioned in the main body of the text, we have used loop QMC simulations as reference data to judge the accuracy of iMPS and iPEPS when simulating the Heisenberg model. For these simulations we have relied on version 4.0a1 of the loop QMC code available as part of the ALPS project.\cite{bauer2011alps} Since loop QMC simulations are carried out both at finite temperature as well as finite size, obtaining ground state estimates in the thermodynamic limit will in general require finite-T as well as finite-size extrapolations. In order to simplify the procedure we have performed all simulations at temperatures low enough to render the finite T variations comparable to the statistical error. We found that temperatures in the range $T \sim [0.01,0.003]$, depending on cylinder width, were enough to obtain negligible variations.

To obtain energy estimates in the infinite-length limit, we carried out simulations on systems of various lengths $L \in [32,1024]$. This data was then extrapolated assuming a scaling of the form $E(L) = a + b (1/L)$ which, given that the system always has a non vanishing energy gap, is a reasonable assumption. For the actual energy estimates we obtained, see Table \ref{eq:lqmc-heisenberg-data} in Appendix \ref{appendix:numerical_data}.

\section{iPEPS Simulations} \label{appendix:ipeps}

\begin{figure}[!h]
	\centering
	\includegraphics[trim={7.5cm 2cm 0cm 0cm},clip,width=0.8\textwidth]{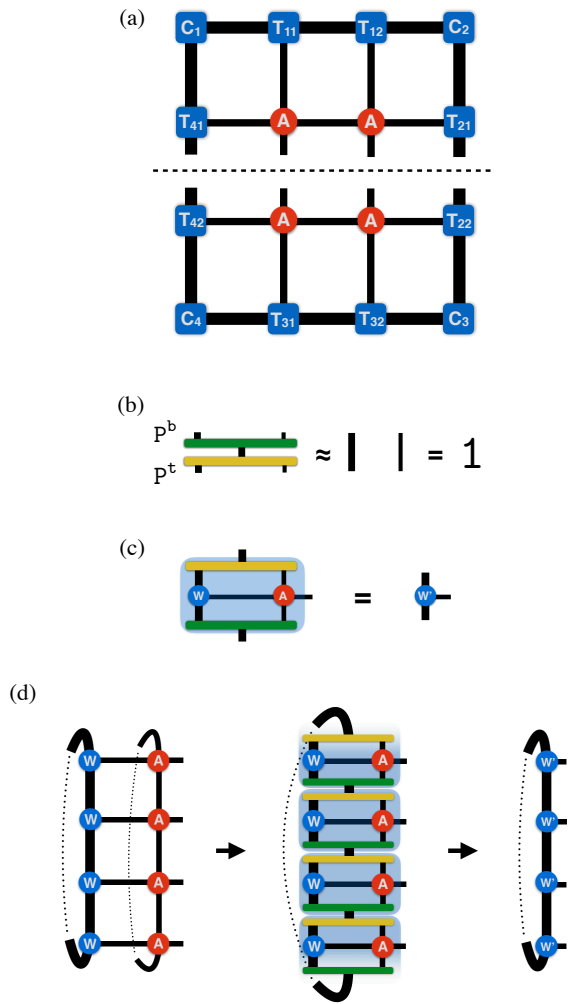}
	\caption{Summary of iPEPS contraction scheme on a cylinder. (a) Tensor network employed for the computation of $P^b$ and $P^t$ projectors on the infinite plane via the CTM algorithm.\cite{corboz2014competing} (b) Approximate resolution of identity in terms of tensors $P^b$ and $P^t$, \ie $P^t \cdot P^b \approx 1 $. (c) Definition of the compressed boundary tensor $W'$. (d) Contraction and compression of a column of PEPS tensors. }
	\label{fig:compression}
\end{figure}

In this section we provide a brief summary of the procedure employed to obtain the iPEPS data presented in the main text.

\begin{figure}
	\includegraphics[trim={0cm 2cm 0cm 2.5cm},clip,width=0.5\textwidth]{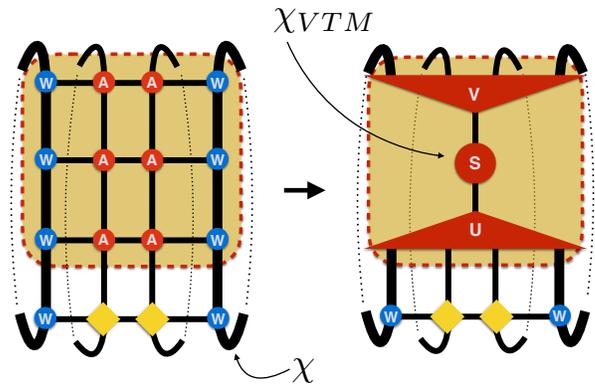}
	\caption{Compression of the cylindrical environment in the measurement of a 2-site observable (diamond-shaped yellow tensors).}\label{fig:cylinder_compression}
\end{figure}

For the purposes of this paper we can reduce the iPEPS procedure on cylinders to a few steps, which we illustrate in Fig.~\ref{fig:compression} (the diagrams show translation invariant systems only for the sake of simplifying the illustrations). The first involves obtaining effective representations of semi-infinite cylinders, which one may achieve via the boundary tensors $W$. We refer to the number of states in the auxiliary (vertical) indices of W as $\chi$, see Fig.~\ref{fig:compression}(d) and Fig.~\ref{fig:cylinder_compression}. These tensors can be obtained by the iterative absorption of full columns of PEPS tensors onto some initial set of $W$ tensors. We find that recycling the boundaries obtained from a simulation on the infinite plane provides good starting points for this construction.

To prevent the boundary bond dimension $\chi$ from growing indefinitely as columns of PEPS tensors are absorbed, we employ approximate resolutions of identity $P^t \cdot P^b \approx 1$, in terms of projectors $P^t$ and $P^b$, see Figs. \ref{fig:compression}(b) and \ref{fig:compression}(c).  The computation of these projectors can be done in a number of different ways. Here we also find that recycling the projectors obtained from a CTM formulation\cite{corboz2014competing} on the infinite plane can be used to give good results. We have carried out simulations using alternative variants and have found results which are consistent, \ie all results agree within the error bars we quote below.

When computing the energies, a cylinder made up of the boundary tensors $W$ and up to 2 columns of PEPS tensors must be contracted. Here we find that employing an SVD compression on the cylindrical environment surrounding the measurement tensors, see Fig.~\ref{fig:cylinder_compression}, provides a controlled way of improving the efficiency of the algorithm at the expense of introducing an additional parameter $\chi_{VTM}$ corresponding to the number of singular values kept. For the examples considered here we found that the number of singular values $\chi_{VTM}$ required for convergence of the energies was reduced quite rapidly as the width of the cylinders was increased. This is a simple consequence of the fact that the spectrum of this vertical transfer matrix thins down exponentially fast with the cylinder-to-unit-cell-width ratio. Also, in order to improve efficiency we have employed, for the larger values of D, an additional compression step which reduces the vertical transfer matrix inside the brown box in Fig.~\ref{fig:cylinder_compression} from a 4-column object to a 2-column object. This compression happens in a manner completely analogous to the procedure illustrated in Fig.~\ref{fig:compression}. An additional parameter $\chi'$ was introduced for this compression and its proper convergence also monitored.

\begin{figure}[h]
	\begin{tabular}{c}
		\includegraphics[trim={0cm 0cm 0cm 1cm},clip,width=0.5\textwidth]{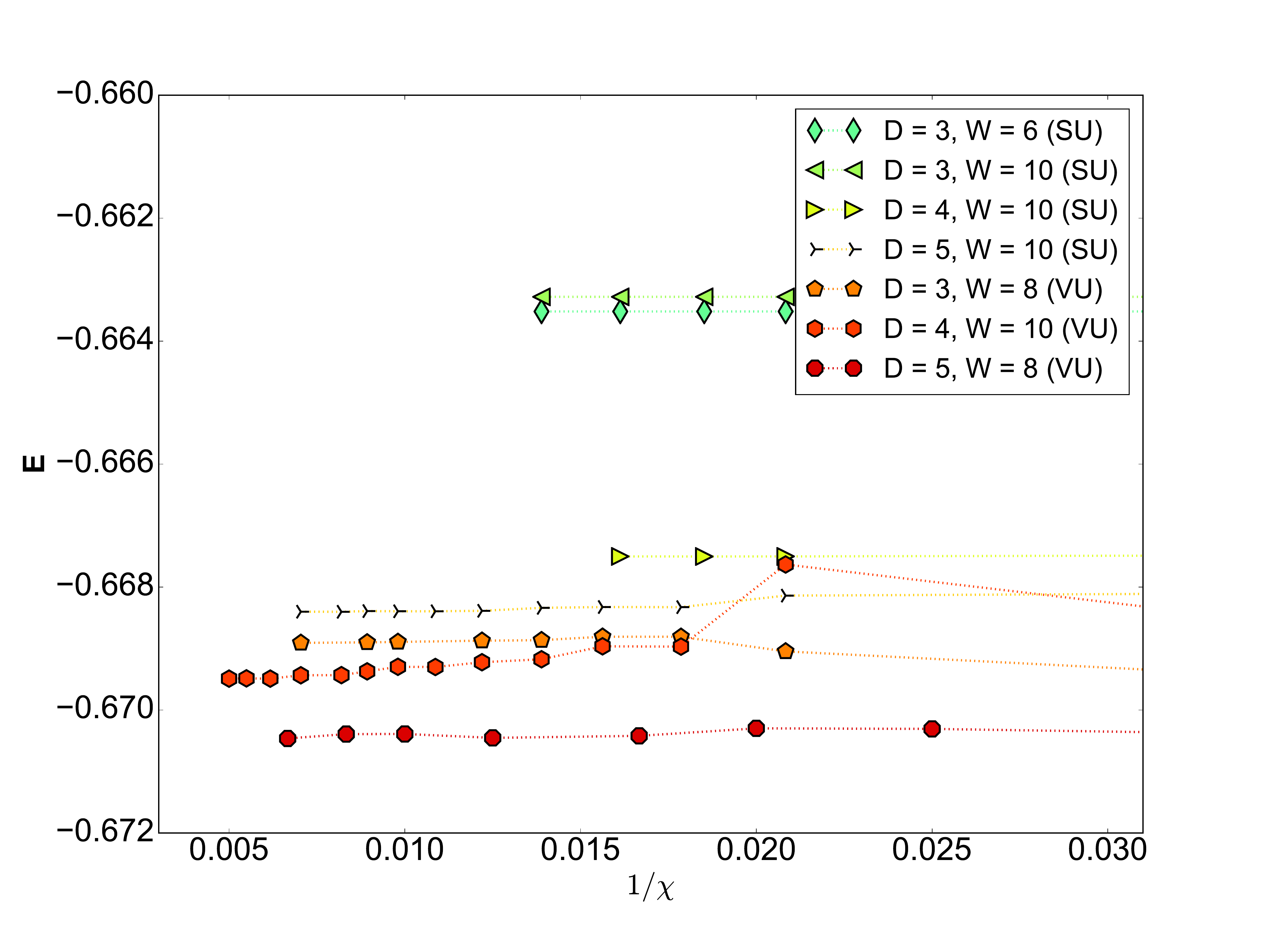}
		\\
		\includegraphics[trim={0cm 0cm 0cm 0cm},clip,width=0.5\textwidth]{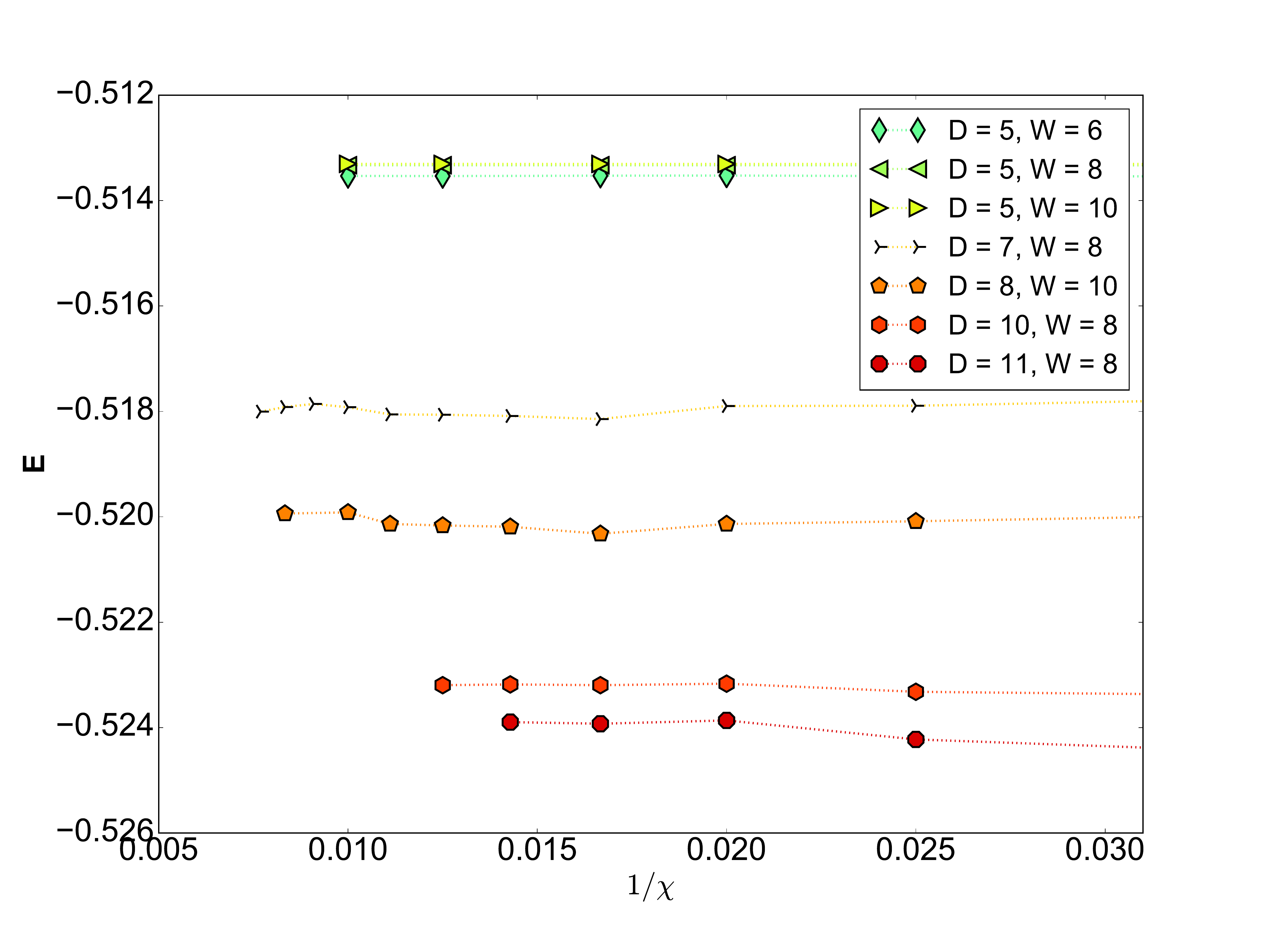}
	\end{tabular}
	\caption{Convergence of iPEPS energies as a function of the inverse boundary bond dimension $\chi$ mentioned in Appendix \ref{appendix:ipeps}. Top: sample convergence data for the Heisenberg model. For this data a value of $\chi_{VTM}$ large enough to exhibit negligible variations was chosen. Bottom: sample convergence data for the half-filled Hubbard model. Here an approach in which $\chi_{VTM}$ was scaled proportionally to $\chi^2$ was used.}\label{fig:convergence}
\end{figure}

Simulations for the Heisenberg model were carried out both with and without preserving the S$_z^{total}$ U(1) symmetry of the model, where the non-symmetric data was used to generate Fig.~\ref{fig:heisenberg-imps-vs-ipeps-var-params} in the main text. We present our energy estimates for the Heisenberg model shown in the main text using the simple and variational updates in tables \ref{eq:ipeps-su-heisenberg-data} and \ref{eq:ipeps-vu-heisenberg-data} of Appendix \ref{appendix:numerical_data}, respectively. We estimate the error bars of the data based on the convergence of the energies as a function of all auxiliary bond dimensions. For the simple update we estimate error bars to be smaller than $\pm 0.0001$ for all values of D. Error bars in the variational update are estimated to be at or below $\pm 0.0003$ for all values of D. In Fig.~\ref{fig:convergence} we provide some sample data illustrating the convergence behavior we found as a function of the auxiliary bond dimension $\chi$.

Simulations for the Hubbard model were carried out preserving both U(1) quantum numbers associated to S$_z^{total}$ and charge conservation. We present our variational energy estimates for the Hubbard model shown in the main text in table \ref{eq:ipeps-vu-Hubbard-data} of Appendix \ref{appendix:numerical_data}. We estimate error bars of this data to be smaller than $\pm 0.0001$ for $D = 5$ and around $\pm 0.0003$ for $D > 5$.

We have constrained our simulations to use checkerboard unit cells, \ie $2\mathbf{x}2$ unit cells with only two types of tensors. All simulations were carried out using real double-precision arithmetic.

\section{iMPS Simulations} \label{appendix:imps}
All iMPS simulations were carried out using an implementation of the iDMRG algorithm based on the ALPS libraries, which we hope to make public soon. Simulations for the Heisenberg model were carried out both with and without preserving the U(1) symmetry corresponding to the conservation of S$_z^{total}$ present in the model. The symmetric and non-symmetric simulations were used to generate Fig.~\ref{fig:heisenberg-imps-vs-ipeps-width} and Fig.~\ref{fig:heisenberg-imps-vs-ipeps-var-params} in the main text, respectively. We present our U(1) symmetric variational energy estimates for the Heisenberg model shown in the main text in table \ref{eq:imps-heisenberg-data} of Appendix \ref{appendix:numerical_data}. Similarly, all simulations for the Hubbard model were carried out preserving both U(1) quantum numbers associated to S$_z^{total}$ and charge conservation. We present our variational energy estimates for the Hubbard model shown in the main text in table \ref{eq:imps-Hubbard-data} of Appendix \ref{appendix:numerical_data}.

We have restricted the size of the optimization unit cells to be twice the width of the cylinder and used a bottom-to-top left-to-right zig-zag pattern to cover the system. All simulations were carried out using real double-precision arithmetic.

\section{Numerical Data} \label{appendix:numerical_data}

In this section we provide tables containing all the numerical data presented in the main text.

\subsection{Heisenberg Model}

\begingroup
\squeezetable

\begin{table}[h!]
	\begin{tabular}{|c||c|}
		\hline
		Width & $E_0$ \\ \hline \hline
		4 & -0.683282(2) \\ \hline
		6 & -0.672788(1) \\ \hline
		8 & -0.670760(2) \\ \hline
		10 & -0.670101(2) \\ \hline
		12 & -0.669815(2) \\ \hline
		14 & -0.669677(1) \\ \hline
		16 & -0.669594(3) \\ \hline
		\end{tabular}\par
		\caption{Loop QMC estimates for the infinite-length finite-width ground state energies of the $S=1/2$ Heisenberg model on square lattice cylinders.}\label{eq:lqmc-heisenberg-data}
	\end{table}
\endgroup

\pagebreak

\begingroup
\squeezetable

	\begin{table}[h!]
		\begin{tabular}{|c||c|c|c|c|c|c|c|}
			\hline
			D \textbackslash Width &        4  &        6  &        8  &        10 &        12 &        14 &        16 \\
			\hline \hline
			2 & -0.6624 & -0.6595 & -0.6594 & -0.6594 & -0.6594 & -0.6594 & -0.6594 \\ \hline
			3 & -0.6678 & -0.6635 & -0.6633 & -0.6633 & -0.6633 & -0.6633 & -0.6633 \\ \hline
			4 & -0.6744 & -0.6685 & -0.6677 & -0.6675 & -0.6675 & -0.6675 & -0.6675 \\ \hline
			5 & -0.6749 & -0.6698 & -0.6686 & -0.6683 & -0.6683 & -0.6683 & -0.6682 \\ \hline
			\end{tabular}\par
			\caption{iPEPS estimates for the infinite-length finite-width ground state energies of the $S=1/2$ Heisenberg model on square lattice cylinders using tensors optimized with the simple update. Error bars are estimated to be at or below $\pm 0.0001$.}\label{eq:ipeps-su-heisenberg-data}
	\end{table}
\endgroup

\begingroup
\squeezetable
	\begin{table}[h!]
		\begin{tabular}{|c||c|c|c|c|c|c|c|}
				\hline
				D \textbackslash Width &        4  &        6  &        8  &        10 &        12 &        14 &        16 \\
				\hline \hline
				2 & -0.6681 & -0.6632 & -0.6626 & -0.6625 & -0.6625 & -0.6625 & -0.6625 \\ \hline
				3 & -0.6743 & -0.6700 & -0.6689 & -0.6686 & -0.6684 & -0.6684 & -0.6683 \\ \hline
				4 & -0.6729 & -0.6709 & -0.6699 & -0.6695 & -0.6693 & -0.6692 & -0.6692 \\ \hline
				5 &       	& -0.6715 & -0.6704 & -0.6699 & -0.6696 & -0.6695 & -0.6695 \\ \hline
				\end{tabular}\par
				\caption{iPEPS estimates for the infinite-length finite-width ground state energies of the $S=1/2$ Heisenberg model on square lattice cylinders using tensors optimized variationally on the infinite plane. Error bars are estimated to be at or below $\pm 0.0003$. }\label{eq:ipeps-vu-heisenberg-data}
	\end{table}
\endgroup

\begingroup
\squeezetable
	\begin{table}[h!]
		\begin{tabular}{|c||c|c|c|c|c|c|c|}
			\hline
					m \textbackslash Width &   4 &   6 &   8 &  10 &  12 &  14 &  16 \\
					\hline \hline
					550  & -0.68328 & -0.67273 & -0.67042 &  -0.66895 &  -0.66704 &  -0.66456 &  -0.66173 \\ \hline
					1050 & -0.68328 & -0.67278 & -0.67047 &  -0.66959 &  -0.66841 &  -0.66675 &  -0.66463 \\ \hline
					2050 & -0.68328 & -0.67279 & -0.67067 &  -0.66988 &  -0.66915 &  -0.66811 &  -0.66667 \\ \hline
					4096 & -0.68328 & -0.67279 & -0.67074 &  -0.67001 &  -0.66952 &  -0.66890 &  -0.66797 \\ \hline
		\end{tabular}\par
		\caption{iMPS estimates for the infinite-length finite-width ground state energies of the $S=1/2$ Heisenberg model on  square lattice cylinders.}\label{eq:imps-heisenberg-data}
	\end{table}
\endgroup

\pagebreak

\subsection{Hubbard Model}

\begingroup
\squeezetable
\begin{table}[h]
	\begin{tabular}{|c||c|c|c|}
						\hline
						D \textbackslash Width &   6 &   8 &  10 \\
						\hline \hline
						5  & -0.5135 & -0.5133 & -0.5133 \\ \hline
						7  & -0.5182 & -0.5179 & -0.5179 \\ \hline
						8  & -0.5207 & -0.5202 & -0.5199 \\ \hline
						10 & -0.5234 & -0.5232 & -0.5231 \\ \hline
						11 & -0.5242 & -0.5239 & -0.5238 \\ \hline
	\end{tabular}
	\caption{iPEPS estimates for the infinite-length finite-width ground state energies of the half-filled Hubbard model at $U/t=8$ on square lattice cylinders using tensors optimized with the full update. Error bars are estimated to be smaller than $\pm 0.0001$ for $D = 5$ and around $\pm 0.0003$ for $D > 5$.}\label{eq:ipeps-vu-Hubbard-data}
\end{table}
\endgroup

\begingroup
\squeezetable
\begin{table}[h]
	\begin{tabular}{|c||c|c|c|c|}
						\hline
						m \textbackslash Width & 6 & 8 & 10 \\
						\hline \hline
						1024 & -0.52248 & -0.51402 &  -0.50221 \\ \hline
						2048 & -0.52411 & -0.51853 &  -0.51059 \\ \hline
						4096 & -0.52491 & -0.52149 &  -0.51520 \\ \hline
						8192 & -0.52524 & -0.52321 &  -0.51896 \\ \hline
	\end{tabular}
	\caption{iMPS estimates for the infinite-length finite-width ground state energies of the half-filled Hubbard model at $U/t=8$ on square lattice cylinders.}\label{eq:imps-Hubbard-data}
\end{table}

\endgroup

\clearpage

\bibliography{imps_vs_ipeps}
\end{document}